\documentclass [12pt]{article}
\begin{document}
\title{Calculation of Heavy Meson Potential Coefficients by Solving Lippman-Schwinger Equation}
\author{ P  Sadeghi  Alavijeh\thanks{ph.parva@yahoo.com}       ,\ M  Monemzadeh\thanks{monem@kashanu.ac.ir}     and  \ N  Tazimi\thanks{nt$_{-} $physics@yahoo.com}      \\\\
\it\small{{Department of Physics, University of Kashan, Kashan, Iran }}}
\date{}
\maketitle
\begin{abstract}

In the present work, we study meson systems consisiting of quark-antiquark. We solve Lippman-Schwinger equation numerically for heavy meson systems . We attempt to find a non-relativistic potential model through which we can solve the quark-antiquark bound state problem. The co-efficients so obtained are in agreement  with Martin potential coefficients.
Via this method we also determine the strong coupling constant for the mesons $t\bar{t}$, $t\bar{c}$ and $t\bar{b}$ which is a coefficient of Cornel potential.
\end{abstract}


\newpage 
\section{Introduction}
The heavy quark potential is an important quantity associated with confinement. This
quantity can be almost accurately calculated via lattice simulations [1]. This potential
is speciﬁed through a non-relativistic effective  theory, yet it also deserves serious investigation in perturbative QCD. Lattice QCD simulations are used for analyzing at long
distances, but for short distances, perturbation theory works better. \\

There exists a variety of experimental data on light and heavy mesons, and it is still
increasing \cite{2}. When the heavy quark mass ($ m_{Q}$) remarkably exceeds  $ \Lambda_{QCD}$ (the QCD scale), the running coupling $\alpha_{s}(m_{Q}) $  is  small. This means that at this scale of the order of    the Compton wavelengh $\lambda\sim 1/ m_{Q} $, perturbative QCD can serve to describe hadrons. 
In the  non-relativistic limit, it has been shown that the interaction between the two  $\bar{Q}$ 
and $Q$ states can be described by a local potential  $ V(r)$, where $ r$  is the relative coordinate
between $Q$ and $\bar{Q}$ (spin is ignored for the moment) \cite{1}. \\

 We  exploit Lippman-Schwinger equation to analyze the quark- antiquark bound state. This equation has recently been used in some studies of  non-relativistic  bound state \cite{3,4,5}. We have investigated some potential models for heavy mesons. We have also identified the stability intervals of some suitable potentials for these systems and obtained
the mass spectrum of these mesons. In our recent work \cite{6}, we have studied the tetraquark
and solved the two-body problem for tetraquark systems of diquark-antidiquark. \\

Having studied some types of appropriate potentials such as Martin \cite{7} and Cornel \cite{8,9}, we found Martin potential as the best one for heavy mesons. This is because of the
potential's larger stability interval \cite{3}. In the present work, we calculate the coefficients
of Cornell and Martin potentials (which are suitable for describing heavy mesons) through
numerical solution of Lippman-Schwinger equation and obtain the physical eigenvalue. It
must be notified  that we have used input parameters diﬀerent from those of ref \cite{3} (quark
mass, binding energy, and potential). In ref \cite{3}, we utilized a revised potential, but we
have made use of Cornell potential in the present paper. \\

It is worth mentioning that in our previous work \cite{3}, we detected appropriate potentials
through this method and obtained $\lambda=1$  within an extensive interval of $r$; therefore, we
have been certain about the appropriateness of Cornell and Martin potentials for heavy mesons in our method. That is, we took the binding energy as already identified and solved Schrodinger equation to find two potential coefficients for each meson.
 
 The procedure of the study is given in sec 1. We explain how to find Martin potential
coefficients in sec 2. In sec 3, we have also used Coulomb potential confinement to predict
the strong coupling constant  for $t\bar{t}$, $t\bar{c}$ and $t\bar{b}$. The results and discussion appear in sec. 4 and finally, sec. 5 gives a concise conclusion of the study.\\

\section{Method}
In this section, we obtain the coefficients of a suitable potential for heavy meson systems by using heavy mesons' binding energies and by solving Lippman-Schwinger equation numerically.\\

The two-body bound state of Schrodinger equation with potential $V$ is:\\
\begin{equation} 
\mid\psi_{b}>=G_{0}V\mid\psi_{b}> 
\label{1}
\end {equation}
where $G_{0} $ is the propagator of a free particle. In configuration space, it is shown as:
\begin{equation} 
\psi_{b}(r)=-m \frac{1}{4\pi}  {\int_{0}^{\infty}}{ dr^\prime} {r^\prime}^{2}{\int_{-1}^{1}} dx^\prime {\int_0}^{2\pi} d\phi ^\prime \frac{exp(-\sqrt{m\vert E_{b}\vert} \vert{r-r^\prime}\vert)}{\vert{r-r^\prime}\vert} V(r^\prime){\psi_{b}(r^\prime)} 
\label{2}
\end {equation}

\begin{equation}
{\psi_{b}(r)}=\int_{0}^ {\infty}dr^{'} \int_{-1}^ {1} dx^{'} M(r,r^{'}, x^{'}) {\psi_{b}(r^{'})} 
\label{3}
\end{equation}
where:
\begin{equation} 
M(r,r^{'},x^{'})= -\frac{m}{2}  \frac{exp((-\sqrt{m\vert E_{b}\vert) }\sqrt{r^{2o}+\acute{r}^{2}-2r \acute{r} \acute{x}})}{\sqrt{r^{2}+\acute{r}^{2}-2r\acute{r} \acute{x}}}{} \\ 
{ V(\acute{r}^{2})} 
\label{4}
\end{equation}
$E_{b}$ stands for the binding energy of two-body bound system. The eigenvalue form of equation (\ref{3}) is:
\begin{equation}
K(E_{b})\vert\psi_{b}>=\lambda(E_{b})\vert\psi_{b}>
\label{5}
\end{equation}
To solve this equation, we use  a Fortran code. It is used to calculate the eigenvalues for a real non-symmetric square matrix. The magnitude of eigenvalue $\lambda$ depends on energy. The energy that results in $\lambda=1$ is the binding energy.  $\lambda=1$ is the highest positive eigenvalues.  The eigenvalue equation  can be solved by direct iteration  method \cite{10}. To discretize  it, the integral Gauss-legendre method \cite{11} is used. It is worth mentioning that to get better results out of the calculations, we consider Gaussian quadrature points for $r$, $ r^{'}$,  and $x^{'}$ as 100 (The more the points, the more accurate the results, although this lowers the running speed of the program). \\
The required input for the program includes system mass, $E _{b}$, potential coefficients, and r-cutoff. 
R-cutoff, which is of the same order of magnitude as the meson radius, is supposed
to be the point where the potential tends to zero. $E_{b}$ is obtained via solving an eigenvalue equation by using quark-antiquark interaction (spin-spin interaction in the potential and spin splitting are ignored). The masses of constituent quarks (MeV) are fitted as: \\
$$m_{c}=1800, m_{b}=5174, m_{t}=174000       $$
Ref \cite{12} also makes use of the same inputs for Martin potential. Table 1 shows heavy meson
masses and heavy meson binding energies. Binding energy is deﬁned as the energy used
when breaking a meson into its components, i.e.quark and antiquark, so it is negative \footnote{In order to separate the quark from the antiquark, huge energy of the order of several MeV is required,
i.e. the rest mass energy of the quarks in free state exceeds their total energy while they are inside the
meson. The energy required to separate the quarks is the binding energy of the meson.}. 
As we know, meson mass is calculated through equation (6)\\
\begin{equation}
M(meson)=m_{quark}+m_{antiquark}+E_{b}
\end{equation}

\begin {table}[h]
\centering
\caption {Heavy meson mass and binding energies  (MeV) }
\begin{tabular} { c   c  c  c } 
\hline Spin & {$\quad Meson$} &{ Meson Mass (PDG[13])   }&{$ E_{b} $}  \\ \hline
S=1 & {$\quad c \bar{c}  (J/\psi)$} &{$ \quad 3096.9 $} &{$ \quad -503.1 $}  \\ \hline
S=0 &{$\quad c\bar{c}(\eta_{c}) $} &{$ \quad  2979.9$} &{$\quad -620.1$}\\  \hline
S=1 &{$\quad b \bar{b} (\Upsilon) $} &{$ \quad9460 $}& {$\quad- 888$}  \\ \hline 
S=0 &{$\quad b \bar{b} (\eta_{b}) $}  &{$ \quad 9398 $}&{$\quad -850 $} \\ \hline 
\end{tabular}
\label{1}
\end{table}

$E_{b}$ in table (1) is obtained from equation (6). Based on the information presented in
table 1, we introduce the reduced mass of the meson's constituent quarks and also the
energies into the program [3]. First, we place the potential in the program:
\begin{equation}
 V(r)=(ra_{n}/ \hbar c)^{z}
  \label{25}
  \end{equation} 
  where 
 $a_{n}=1000 MeV$ is an input .We have introduced the coefficient $a _{n} $ to correct the dimension so that the dimensionof each term of the potential is given in MeV.
  The parameter $\hbar c$ is included in the equation for dimension coordination. $V $ is expressed in terms of  $MeV $ in the program and $ r$ in terms of  $fm$. In
this part, we assigned $z$ an arbitrary quantity. The purpose was to obtain $\lambda=1$, so $z$ was assigned lower or higher values. We carried out this process of iteration until $\lambda=1$ was
obtained as the program output. There is a direct relationship between the coefficient
$z$ and the program output, i.e. the $\lambda$ spectrum. The higher the coefficient, the higher
the values obtained in the $\lambda$ spectrum. The size and direction of each iteration
depends on what $\lambda$ spectrum is obtained following each iteration. If the values in the
spectrum are much larger/smaller than 1, the next assigned $z$ should naturally be much
larger/smaller than the previous $z$. But when we obtain a $\lambda$ spectrum very near to 1, the
next assigned $z$ will be very near to the previous one. Of course, the rate of accuracy set
in this stage of the study is 0.01, then $0.99 \leq \lambda \leq 1.01$ is acceptable. Thus, the quantities
that yielded eigenvalues either greater than 1.01 or smaller than 0.99 were ignored. The
results of running the program show that at $z$ = 0.1, the desired output is obtained. \\

In the next stage, we introduce the coefficient $k$ and take the potential as $k(r^{0.1})$. We
then follow the above procedure, i.e. we obtain $\lambda=1$ by assigning different  values to $k$.
The rate of accuracy; however, is 0.001. Therefore, any $0.999 \leq \lambda \leq 1.001$ is acceptable.
Finally, we introduce the constant $d$ and increase the accuracy up to 0.0001 . In fact, we
don't introduce the potential at once but rather stage by stage. Initially, the coefficient
$z$ is introduced and the predetermined acceptable error is 0.01. Later as we introduce
the other coefficients, we approach the full form of the potential and the tolerance is
enhanced to 0.01 for $k$ and finally to 0.0001 for $d$.
\begin{equation}
V(r)=k(a_{n }r)^{z}+d
  \label{25}
  \end{equation}
  
We are using tables 1 and 2 as input. Table 2 illustrates the results. The cells marked with
an asterisk $(*) $ contain results that comply with the coefficients of Martin potential. In
this potential, $r$ is defined in terms of $GeV^{-1}$. The cells containing a dash $(-) $ indicate
potential coefficients which didn't enjoy the desired accuracy. Through the procedure
already described, we have managed to obtain values of $z$, $k$, and $d$ which are signiﬁcant
for different  potentials. Our results are spin-dependent because the potential models we
are considering are spin-dependent.
\newpage 
\begin {table}[h]
\centering
\caption{Potential  exponent and coefficients obtained from the program}
\begin{center}
\begin{tabular}{cccc}
\hline
Meson & z &k(MeV)&d(MeV)\\
\hline
 & 0.0931&6886&- \\ 
c\={c}(S=1) & &6890&-\\ 
 & 0.0972*&6899*&8071* \\ 
 & &6910&-\\
\hline
 & 0.0981*&6990&- \\ 
 c\={c}(S=0)& &6930*&8980*\\ 
 & &&8591 \\ 
 & 0.1*&6849&6849\\ 
 & & &6862\\  
 & &6862 &-\\ 
\hline
b\={b}(S=0) &0.0983*&6883*&8669 *\\ 
 & &&9465\\ 
\hline
 Martin potential coefficients \cite{7} & 0.1&6898&8093\\
 \hline
\end{tabular}
\end{center} 
\end{table}
\section{Calculating Heavy Meson Strong Coupling Constant}
 The running coupling in QCD is small at high energies, while it increases to order 1 at low
energies. This feature is asymptotic freedom of QCD. Mesons containing top quark are
unstable because of their short lifetime. However, within this short lifetime, the strong
coupling constant can be anticipated (we are aware that this application may be purely
academic because of the non-existence of the corresponding bound states).\\

  In this section, our anticipated  strong coupling constant for $t\bar{t}$,  $t\bar{c}$ and $t\bar{b}$  are presented, thus here we use Cornell potential. Cornell potential is composed of two terms. One is Coloumb potential used for interaction in small distances i.e. $V\propto-\alpha_{s}/r $ at $r\rightarrow0$.  The
other is a linear potential capable of describing conﬁnement. It is used for large distances,
i.e. $V\propto r $ at large $r$. This potential has absorbed much attention in hadron physics and
is definitely of strong interaction type. This potential is of the following form  \cite {8,9}:
\begin{equation}
V(r)=fr-\frac{4}{3}  \frac{\alpha_{s}\hbar c}{r} 
  \label{7}
  \end{equation} 
where  $ f= 150000  MeV^{2}$ \cite {15} is the string tension .   Here, $\hbar c$ is also used for dimension coordination. $\alpha _{s}$ is the coefficient to be worked out.
Now, we introduce the energy (table  (\ref{3})) and potential (eq. 9)  into the program.  We assign an arbitrary initial value  to $\alpha _{s}$ (we assigned 0.01) and then
search through the immediate neighbourhood iteratively with a step-size of 0.0001   until $\lambda=1\pm0.01$ is encountered. The results are presented in table  (\ref{5}). 
Varying $\alpha _{s}$  leads to
diﬀerent $\lambda$ spectra. However, we seek to arrive at results (i.e. spectrum) which are closest to 1. In other words, accuracy depends on the minimum allowed error of calculation
outcome. It is worth mentioning that strong coupling constant for  $c\bar{c}$ and $b\bar{b}$ mesons can also be calculated by this method.\\

We are using table 3 as input. Our results are available in tables 4 and 5. According
to \cite{14}, coupling constant is inversely correlated with mass, i.e.\\
$$  m_{t\overline{c}}\leq m_{t\overline{b}}\leq m_{t\overline{t}}\Rightarrow \alpha_{t\overline{t}}\leq \alpha_{t\overline{b}}\leq \alpha_{t\overline{c}}$$
Our results in table 4 also indicate this inverse correlation. We used this method to identify the coupling constant of $c\overline{c}$ and $b\overline{b}$ mesons. The $\alpha_{s}$ we obtained is similar to the $\alpha_{s}$ in \cite{16}. The results appear in table 5.\\

As stated in the introduction, we have earlier investigated certain potential models
appropriate for heavy mesons to calculate the binding energy of these systems. Since this
method is potential-independent, the obtained binding energy applies to other potentials
too \cite{17}. In \cite{3}, we arrived at the binding energy of $t\overline{t}$ and $t\overline{c}$ systems through a revised
potential (we have also calculated the binding energy of $t\overline{b}$ by means of this potential).
In this paper, we used that binding energy as the input of the program and obtained
potential coefficients diﬀerent from the previous work. As two examples, we obtained
Martin potential coefficients for $b\overline{b}$ and $c\overline{c}$ systems and obtained the coupling constant in
Cornell potential for $t\overline{b}$, $t\overline{c}$, $t\overline{t}$, $b\overline{b}$ and $c\overline{c}$ systems. However, we could have alternatively
calculated Martin potential coefficients for t-quark mesons.
\begin {table}[h]
\centering
\caption{Introduced energies for heavy mesons \ref{3}}
\begin {center}
\begin{tabular}{ccc}
\hline Meson &S  &E(MeV)  \\ \hline
$ t \bar{t} $ & 1 & -11.296 \\ 
$ t \bar{t} $  & 0 &-36.948  \\ 
$ t \bar{b} $  & 0 &  -195.0870\\ 
 $ t \bar{c} $& 0 &-550  \\ 
$ t \bar{b} $  & 1 &-641.7242  \\ 
\hline 
\end{tabular}
\end{center}
\end{table}
\begin {table}[h]
\centering
\caption{Strong coupling constants  obtained for $t \bar{t}$, $ t \bar{b}$, and $ t \bar{c} $   mesons}
\begin {center}
\begin{tabular}{cccccc}
\hline 
Meson & $ t \bar{t}( S=0) $  &$ t \bar{t}( S=1) $  &$  t \bar{b}( S=0) $ & $t \bar{b}( S=1) $ &   $ t \bar{c}( S=0) $   \\ 
\hline
 $\alpha _{s}$   &0.2340  & 0.1998&  0.3240&0.3060&0.4260   \\ 
\hline 
\end{tabular}  
\end{center}
\end{table}

\section{Results and Discussion}
We made use of mass, r-cutoff, and binding energy of each meson to solve Schrodinger
equation and diagonalize the kernel. We searched iteratively through parameter space to
obtain coefficients for proposed potential model until desired eigenvalues were met within
acceptable tolerance. We found exclusive potential coefficients for each meson. The pro-
cedure led to the identiﬁcation of two appropriate potential coefficients for heavy mesons.\\

It must be notified  that the two potentials used in this paper are spin-independent.
The coefficient diﬀerences observed across $S=0$ and $S=1$ states result from the different binding energies that we used for these two states (tables 1 and 3). Thus, spin
indirectly affects the mass and coefficients of the potential.\\

\begin {table}[h]
\centering
\caption{Strong coupling constants  obtained for $c \bar{c}$ and $ b \bar{b}$    mesons}
\begin {center}
\begin{tabular}{cccccc}
\hline 
Meson & $ \alpha_{s} $ in ref[16]  & Calculated $ \alpha_{s} $   \\ 
\hline
 $c \bar{c}$   &0.51  & 0.48   \\ 
  $b \bar{b}$   &0.33  & 0.31   \\ 
\hline 
\end{tabular}  
\end{center}
\end{table}

\section{Conclusions} 
In this study, we managed to calculate the coefficients of one potential through solving
Lippman-Schwinger equation numerically and obtain the physical eigenvalue. This po-
tential's coefficients are very close to Martin potential coefficients. Additionally, since
we arrived at acceptable results in the program, we managed to obtain $\alpha _{s}$ considered as
an unidentified coefficient in the Cornell potential for $t \bar{t}$, $ t \bar{b}$, $ t \bar{c}$, $ b \bar{b}$ and $ c \bar{c} $  mesons. The
method practiced, therefore, is a good one to identify the potential coefficients used for
heavy mesons.\\
\section{Acknowledgments }
We are pleased to thank  the University of Kashan for Grant No. 65500.4\\

\end{document}